\documentclass{ws-procs9x6-cpt22}
\begin{document}

\def\etal {{\it et al.}}
\def\cO{\mathcal O}
\def\cL{\mathcal L}
\def\vev#1{\langle {#1}\rangle}
\def\kv{\vev{k}}
\def\kb{\overline{k}}

\title{Concepts in Lorentz and CPT Violation}

\author{V.\ Alan Kosteleck\'y}

\address{Physics Department, Indiana University\\
Bloomington, IN 47405, USA}

\begin{abstract}
This contribution to the CPT'22 meeting provides a brief review  
of some concepts in Lorentz and CPT violation.
\end{abstract}

\bodymatter

\section{Introduction}

In recent years,
substantial advances have been made 
in the theory and phenomenology of Lorentz and CPT violation,
driven by the intriguing possibility
that experimental searches for the associated effects
could uncover tiny observable signals
from an underlying unified theory such as strings.\cite{ksp}
This contribution to the CPT'22 proceedings
summarizes a few concepts in the subject,
focusing on the approach that uses effective field theory\cite{eft}
to construct a general realistic framework 
describing physical effects.\cite{ck,ak04,kl21}

\section{Foundations}

A key property of Minkowski spacetime
is its invariance under Lorentz transformations.
The Lorentz transformations include spatial rotations and velocity boosts,
which together can be viewed as generalized rotations in spacetime.
Theories manifesting isotropy under these spacetime rotations
have Lorentz invariance (LI),
so a theory with Lorentz violation (LV) incorporates 
one or more spacetime anisotropies.
Experiments testing LI seek to identify possible spacetime anisotropies
by comparing physical quantities at different spacetime orientations.
For example,
symmetry under spatial rotations can be explored 
by comparing at different spatial orientations
the ticking rates of two clocks 
or the lengths of two standard rulers.

Our most successful fundamental theories 
describing nongravitational aspects of Nature 
are constructed on Minkowski spacetime.
However,
the Universe contains gravitational interactions,
which cannot be screened and hence are ubiquitous.
Minkowski spacetime is therefore believed to be unphysical in detail,
with our Universe instead involving Riemann spacetime
or perhaps a generalization.
A generic Riemann spacetime lacks global LI.
Instead,
the relevant spacetime symmetries 
are local Lorentz invariance (LLI)
and diffeomorphism invariance (DI).
A theory with LLI is isotropic
under local spacetime rotations about every point,
while one with DI is unchanged by local translations.
Experiments can test these symmetries 
by comparing properties of objects
at different orientations and locations 
in the neighborhood of a spacetime point.

Most experiments and observations 
involve only weak gravitational fields
and so take place in asymptotically flat spacetime,
a limit of Riemann spacetime 
that reduces to Minkowski spacetime for zero gravity.
In asymptotically flat spacetime,
the standard notion of LI 
turns out to be a combination of LLI and DI.\cite{kl21}
Experiments searching for LV 
are therefore in reality sensitive 
to a combination of local Lorentz violation (LLV)
and diffeomorphism violation (DV).
The corresponding observables manifest
a mixture of local spacetime anisotropy
and spacetime-position dependence.

\section{Theory}

Since no compelling evidence for LV exists to date,
a broad-based and model-independent methodology is desirable 
in the search for possible effects.
Any violations are expected to be small corrections
to the known physics of General Relativity (GR) and the Standard Model (SM),
so it is natural to study LV 
using the approach of effective field theory.\cite{eft}
Typically,
integrating over high-energy degrees of freedom in a theory
generates a specific effective field theory applicable at low energies.
In the context of searches for LV,
however,
the approach can instead be used
to study simultaneously a large class of underlying theories
and determine possible observable effects
in a model-independent way.

The realistic coordinate-independent theory 
incorporating general LV 
that is based on GR coupled to the SM
is known as the Standard-Model Extension (SME).\cite{ck,ak04}
In a realistic effective field theory,
CPT violation comes with LV 
both in Minkowski spacetime\cite{ck,owg} 
and in asymptotically flat spacetime,\cite{ak04} 
so the SME also describes CPT violation in a model-independent way.
The Lagrange density $\cL$ of the theory
includes LV operators of any mass dimension $d$,
with the minimal theory defined as the subset 
of operators of renormalizable dimension $d\leq 4$.
Each term in $\cL$ is constructed as an observer-scalar contraction
of a LV operator $\cO(x)$ with
a coupling coefficient $k(x)$ or its derivatives. 
The coefficients are expected to be suppressed either by powers
of a high-energy scale such as the Planck energy
or via a mechanism such as countershading.\cite{akjt}
The explicit forms of all minimal terms\cite{ck,ak04}
and many nonminimal terms\cite{nonmin-sm,kl19,nonmin-gr,kl21}
are known.

\section{Backgrounds}

Any given coefficient $k(x)$ in $\cL$ 
can be viewed as a prescribed background in spacetime,\cite{ak04}
which may arise as a vacuum expectation value of a field.
All indices carried by $k(x)$ 
are contracted with those of the corresponding operator $\cO(x)$,
so $k(x)$ is covariant under observer local Lorentz transformations
and general coordinate transformations.
However,
$k(x)$ is unaffected by particle local Lorentz transformations
and diffeomorphisms,
which act only on dynamical fields.
Covariant and contravariant local indices on $k(x)$
are physically equivalent because the local metric $\eta$ 
is Minkowski and nondynamical.
In contrast,
covariant and contravariant spacetime indices 
can generate physically distinct effects
because the spacetime metric $g$ is dynamical.
Disregarding derivatives of $k(x)$,
a generic term in $\cL$ thus takes the form
$\cL \supset 
k^{\mu\ldots}{}_{\nu\ldots}{}^{a\ldots}(x) 
\cO_{\mu\ldots}{}^{\nu\ldots}{}_{a\ldots}(x)$, 
where spacetime indices are Greek and local indices are Latin.
The term is LLV when $k(x)$ carries a local index
and is DV when $k(x)$ carries a spacetime index
or varies with $x$.

It is physically useful to distinguish two types of backgrounds $k(x)$,
denoted as $\kv$ and $\kb$.
Spontaneous backgrounds $\kv$
arise dynamically from solving equations of motion,
so they satisfy the Euler-Lagrange equations and are thus on shell.
Their dynamical origin means that small fluctuations around $\kv$ can occur,
so additional modes appear in the effective theory\cite{bk05}
including Nambu-Goldstone\cite{ng60} and massive modes.
Explicit backgrounds $\kb$
are externally prescribed and hence nondynamical.
They are unconstrained by Euler-Lagrange equations
and thus can be off shell,
and no dynamical fluctuations occur.
The spontaneous or explicit nature of the LLV and DV
can be used to classify terms in $\cL$ 
and determine their physical implications.\cite{kl21}
The geometry of gravity is affected differently in the two scenarios.
For spontaneous violation,
it can remain Riemann or Riemann-Cartan.\cite{ak04} 
For explicit violation,
in contrast,
the geometry typically cannot be Riemann 
and is conjectured instead to be Finsler,\cite{ak04,rb15,kl21,finsler}
leading to unique gravitational effects.\cite{kl21-2}

\section{Applications}

By virtue of its generality and model independence,
the SME can be expected to contain the large-distance limit
of any realistic theory with LV.
The background coefficients affect
the behavior of experimentally known force and matter fields
and hence can be used to predict possible observable signals 
of Lorentz and CPT violation.
The physical definition of a given particle species 
can be affected,\cite{kps22} 
and both its free propagation and interactions 
can be modified in a flavor-dependent way.
The effects can depend on the magnitude and orientation of momenta and spins
and can differ between particles and antiparticles.
For spontaneous LV,
the dynamical fluctuations around $\kv$ can also modify the physics
and can even play the role of the photon or graviton.\cite{bk05}
In any scenario,
the coefficients are the targets for experimental searches.
Since a coefficient changes under observer frame transformations,
the inertial frame used to present results must be specified.
In practice no laboratory is inertial,
and the Earth's rotation and revolution imply
that LV measurements typically exhibit time variations.\cite{ak98}
Instead, 
the canonical choice to report experimental results 
is the Sun-centered frame.\cite{scf}
Many experiments have achieved impressive sensitivities to coefficients 
expressed in this frame.\cite{tables}

The SME framework also has applications in broader contexts.
One concerns searches for new LI physics beyond GR and the SM.
By virtue of its inclusion of all effective background couplings,
the framework can describe physical effects 
from any new field producing a background 
over a spacetime region of experimental relevance,
and hence it can be used to deduce bounds on new LI physics
from existing constraints on LV.
This technique has led,
for example,
to tight constraints on torsion\cite{torsion}  
and nonmetricity,\cite{nonmetricity}
and similar ideas have been adopted 
for spacetime-varying couplings and ghost-free massive gravity.\cite{other}
Another application in a different context
involves the description of emergent LI 
in certain condensed-matter systems.
For example,
properties of Dirac and Weyl semimetals
are calculable in the SME framework.\cite{semimetal}
Future explorations in these and other contexts
offer excellent prospects for conceptual and practical advances.

\section*{Acknowledgments}

This work was supported in part
by US DoE grant {DE}-SC0010120
and by the Indiana University Center for Spacetime Symmetries.

\end{document}